%% file: 4nnleft.tex
\documentclass[aps,prl,showpacs,twocolumn,preprintnumbers,superscriptaddress,nofootinbib,floatfix,10pt]{revtex4-1}
\usepackage{amsfonts,amssymb,stmaryrd,latexsym,amsmath,braket}
\usepackage{mathrsfs,dsfont}
\usepackage{graphicx,subfigure}
\usepackage{comment}
\usepackage{times}
\usepackage{slashed}
\usepackage{bm}
\usepackage{braket}
\usepackage{color}
\usepackage{longtable}
\usepackage{multirow}
\usepackage{hyperref}
\hypersetup{
    colorlinks=true,
    linkcolor=blue,
    citecolor=blue,
    urlcolor=blue
}
\usepackage{bibunits}
\newcommand{\beginsupplement}{%
        \setcounter{table}{0}
        \renewcommand{\thetable}{S\arabic{table}}%
        \setcounter{figure}{0}
        \renewcommand{\thefigure}{S\arabic{figure}}%
        \setcounter{equation}{0}
        \renewcommand{\theequation}{S\arabic{equation}}%
     }

\makeatletter
\let\saved@includegraphics\includegraphics
\AtBeginDocument{\let\includegraphics\saved@includegraphics}
\makeatother

\graphicspath{{figures/}}
\begin{document}


\title{Searching for the Tetraneutron Resonance on the Lattice}

\author{Linqian Wu}
\affiliation{Peng Huanwu Collaborative Center for Research and Education, International Institute for Interdisciplinary and Frontiers, Beihang University, Beijing 100191, China}
\affiliation{School of Physics, Beihang University, Beijing 102206, China}

\author{Serdar Elhatisari}
\email{selhatisari@gmail.com}
\affiliation{King Fahd University of Petroleum and Minerals (KFUPM), 31261 Dhahran, Saudi Arabia}
\affiliation{Faculty of Natural Sciences and Engineering, Gaziantep Islam Science and Technology University, Gaziantep 27010, Turkey}

\author{Ulf-G. Meißner}
\email{meissner@hiskp.uni-bonn.de}
\affiliation{Helmholtz-Institut f\"ur Strahlen- und Kernphysik and Bethe Center for Theoretical Physics, Universität Bonn, D-53115 Bonn, Germany}
\affiliation{Institute for Advanced Simulation (IAS-4), Forschungszentrum J\"ulich, D-52425 J\"ulich, Germany}
\affiliation{Peng Huanwu Collaborative Center for Research and Education, International Institute for Interdisciplinary and Frontiers,
 Beihang University, Beijing 100191, China}

\author{Shihang Shen}
\email[Corresponding author:]{sshen@buaa.edu.cn}
\affiliation{Peng Huanwu Collaborative Center for Research and Education, International Institute for Interdisciplinary and Frontiers, Beihang University, Beijing 100191, China}
\affiliation{School of Physics, Beihang University, Beijing 102206, China}

\author{Li-Sheng Geng}
\email[Corresponding author:]{lisheng.geng@buaa.edu.cn}
\affiliation{School of Physics, Beihang University, Beijing 102206, China}
\affiliation{Sino-French Carbon Neutrality Research Center, \'Ecole Centrale de P\'ekin/School
 of General Engineering, Beihang University, Beijing 100191, China}
\affiliation{Peng Huanwu Collaborative Center for Research and Education, International Institute for Interdisciplinary and Frontiers, Beihang University, Beijing 100191, China}
\affiliation{Beijing Key Laboratory of Advanced Nuclear Materials and Physics, Beihang University, Beijing 102206, China}
\affiliation{Southern Center for Nuclear-Science Theory (SCNT), Institute of Modern Physics, Chinese Academy of Sciences, Huizhou 516000, China}

\author{Youngman Kim}
\affiliation{Center for Exotic Nuclear Studies, Institute for Basic Science, Daejeon 34126, Korea}

\begin{abstract}
The nature of the tetraneutron ($4n$) system remains a pivotal question in nuclear physics.
We investigate the $4n$ system using nuclear lattice effective field theory in finite volumes with a lattice size up to $L=30$~fm, employing both a high-precision N$^3$LO interaction and a simplified SU(4) symmetric one.
The ground-state energy is found to decrease smoothly with increasing box size, showing no plateau characteristic of a resonance.
We further compute the dineutron-dineutron scattering phase shift using L\"uscher's finite-volume method. At the smallest relative momenta, the extracted $2n$--$2n$ $S$-wave phase shift is small, consistent with a weak interaction in the dilute limit. At intermediate momenta, it exhibits a weak attraction with a peak of approximately $10^\circ$ at relative momentum of 60--84~MeV.
While this structure does not constitute a resonance, the corresponding confined $4n$ energy of 1.7--3.3~MeV lies close to the experimentally observed low-energy peak.
\end{abstract}
\maketitle
\date{today}

\section{Introduction}

Bound nuclear systems conventionally form when protons and neutrons coalesce through the strong interaction. However, the combination of attractive strong interactions and the absence of Coulomb repulsion makes the existence of pure neutron nuclei a plausible concept, a quest that dates back to the early 1960s \cite{ogloblin1989nuclei}. Confirming or ruling out the existence of such neutral nuclei, whether as bound or resonant states, is crucial for advancing our understanding of nuclear physics, the strong interaction, universal fermion properties in the unitary limit~\cite{Elhatisari:2016hui,Braaten:2004rn}, and astrophysical objects~\cite{Matsuki:2024ios,Ivanytskyi:2019ynz}.

Substantial experimental and theoretical efforts have been dedicated to this challenge. In 2002, events characteristic of multineutron clusters were observed in the breakup of $^{14}$Be~\cite{Marques:2001wh}. In 2016, a candidate resonant tetraneutron ($4n$) state with an energy of $0.83\pm 0.65\text{(stat.)}\pm 1.25\text{(sys.)}\,$MeV was reported in the missing-mass spectrum of the $^4$He($^8$He,$^8$Be)$4n$ reaction~\cite{Kisamori:2016jie}. Most notably, a recent high-statistics experiment on the $^8$He($p$,$p^4$He)$4n$ reaction observed a clear peak structure near the threshold at $2.37\pm 0.38\text{(stat.)}\pm 0.44\text{(sys.)}\,$MeV~\cite{Duer:2022ehf}.

While a bound $4n$ nucleus is generally believed to be excluded~\cite{Lazauskas:2022mvq,Marques:2021mqf,Bertulani:2002px}, the interpretation of these signals as a resonance remains highly debated. A variety of theoretical approaches, including Jost function analysis in the complex momentum plane \cite{Sofianos:1997nn}, the hyperspherical harmonic method \cite{Grigorenko:2004}, Faddeev-Yakubovsky equations \cite{Lazauskas:2005ig,Hiyama:2016nwn,Carbonell:2017dfl}, the Gaussian expansion method \cite{Hiyama:2016nwn,Carbonell:2017dfl}, the no-core Gamow shell model and density matrix renormalization group \cite{Fossez:2016dch}, the $4n$ response function \cite{Lazauskas:2017cfi}, Alt, Grassberger, and Sandhas (AGS) equations~\cite{Deltuva:2018xoa,Deltuva:2019mnv}, and adiabatic hyperspherical methods~\cite{Higgins:2020avy,Higgins:2020pbe} converge on the absence of a $4n$ resonance. The low-energy peak observed by~\cite{Duer:2022ehf} has been explained as a consequence of dineutron-dineutron correlations through a reaction model study~\cite{Lazauskas:2022mvq}. Reactions with multineutron final states using a nonrelativistic conformal field theory are investigated in~\cite{Hammer:2021zxb}.
Recent tensor-optimized antisymmetrized molecular dynamics studies suggest that a $4n$ resonance at the nuclear surface may be influenced by core attraction~\cite{Wan:2025itw}.
For further insight into multineutron correlations in light nuclei, see~\cite{Zhang:2025uin}. 

In contrast, several many-body calculations have reported a low-lying $4n$ resonance with energies and widths close to the experimental values. Some of these results are obtained by introducing an external confining potential and extrapolating to the zero-trap limit, e.g., in quantum Monte Carlo~\cite{Pieper:2003dc,Gandolfi:2016bth} and no-core Gamow shell--model studies~\cite{Li:2019pmg}. Other approaches do not employ an explicit trap, such as the single-state harmonic oscillator representation of scattering equations extension of the no-core shell model~\cite{Shirokov:2016ywq}.
However, as discussed in Refs.~\cite{Deltuva:2019ngx,Ishikawa:2020bcs}, threshold effects are critical and must be carefully accounted for in such extrapolations. For a comprehensive review of previous efforts, see~\cite{Marques:2021mqf}.

To further elucidate the nature of the $4n$ signal reported in~\cite{Duer:2022ehf}, additional theoretical investigations are needed. In this work, we study the $4n$ system using nuclear lattice effective field theory (NLEFT)~\cite{Lee:2025req,Lahde:2019npb}, a method that has proven very successful in providing a unified description of nuclear phenomena on a discretized lattice, including scattering~\cite{Elhatisari:2025fyu,Wu:2025fkn,Elhatisari:2016hui,Elhatisari:2015iga}, clustering~\cite{Epelbaum:2011md,Epelbaum:2012qn,Shen:2022bak,Shen:2024qzi}, nuclear matter~\cite{Tong:2025fzv,Tong:2024jvs,Ma:2023ahg,Ren:2023ued,Lu:2019nbg}, and the structure of light and heavy nuclei~\cite{Hildenbrand:2025voq,Ren:2025vpe,Song:2025ofd,Niu:2025uxk,Wang:2025swg,Zhang:2024wfd,Hildenbrand:2024ypw,Meissner:2023cvo,Sarkar:2023qjn,Lu:2021tab,Shen:2021kqr}. We calculate the ground-state energy of $4n$ in finite volumes up to a lattice size of $L\simeq 30$~fm. No external trap is applied, as the finite volume itself provides the necessary confinement for the dilute system.

\section{Formalism}

In nuclear lattice effective field theory, the many-nucleon system is solved on a discretized space with lattice spacing $a$ and $L\times L\times L$ sites.
The ground state $|\Psi\rangle$ is obtained by applying Euclidean time projection to an initial wave function $|\Psi_0\rangle$ that has nonzero overlap with the true ground state,
\begin{equation}\label{eq:proj}
  |\Psi\rangle = \lim_{\tau\to\infty} e^{-H\tau} |\Psi_0\rangle,
\end{equation}
where the Hamiltonian $H$ includes the kinetic energy and nuclear interactions.
We adopt the same $\chi$EFT interaction at N$^3$LO order defined in Ref.~\cite{Elhatisari:2022zrb}.
The two-nucleon interaction is constrained by low-energy neutron-proton scattering phase shifts, while the three-nucleon force is fitted to the binding energies of several light and medium-mass nuclei.
For further details on the interaction and the solution of the many-body system, we refer the reader to Ref.~\cite{Elhatisari:2022zrb}.

We employ periodic boundary conditions.
The finite lattice box confines the nucleons, regardless of whether they form a bound state or a continuum state.
To better capture the dilute nature of the $4n$ system and accelerate the convergence to the ground state in Eq.~(\ref{eq:proj}), we use a plane-wave Slater determinant as the initial wave function.
The single-particle wave function is
\begin{equation}\label{eq:}
  \psi_{\mathbf{k}}(\mathbf{n}) = \exp\left(i\frac{2\pi}{L} \mathbf{k} \cdot \mathbf{n}\right),
\end{equation}
with $\mathbf{n} = (n_x, n_y, n_z)$ is the lattice coordinate vector, with each component ranging from $0$ to $L - 1$, and
$\mathbf{k}$ is the lattice momentum vector.
The initial state $|\Psi_0\rangle$ is constructed as a Slater determinant from the two lowest momentum states, each with spin up and down.
Furthermore, we perform an angular momentum projection onto the $A_1^+$ irreducible representation of the cubic group to project out the $0^+$ state and accelerate the Euclidean time evolution \cite{Johnson:1982yq,Lu:2014xfa}.
The ground state energy at a given projection time $\tau$ is calculated as
\begin{equation}\label{eq:}
  E(\tau) = \frac{ \langle \Psi_0|e^{-H\tau/2} H e^{-H\tau/2} |\Psi_0\rangle }{ \langle \Psi_0|e^{-H\tau}|\Psi_0\rangle }.
\end{equation}

It is well known that the tetraneutron system lacks bound subsystems.
However, when confined in a finite volume, the dineutron becomes quasibound.
Following the method for dimer-dimer scattering on a lattice~\cite{Elhatisari:2016hui}, we compute the $2n$--$2n$ scattering phase shifts using L\"uscher's formula.
It relates the two-body energy levels in a periodic cubic box to the elastic scattering phase shifts \cite{Luscher:1986pf,Luscher:1990ux}:
\begin{equation}
    p \cot \delta(p) = \frac{1}{\pi L} {S(\eta)}, \quad \eta = \left( \frac{pL}{2\pi} \right)^2,
\end{equation}
with $S(\eta)$ the regulated three-dimensional $\zeta$ function
\begin{equation}
    S(\eta) = \lim_{\Lambda\to \infty} \left[ \sum_{\vec{\bm{n}},\, |\vec{\bm{n}}| \leq \Lambda} \frac{1}{\vec{\bm{n}}^2 - \eta} - 4\pi \Lambda \right].
\end{equation}
The sum runs over all integer vectors $\mathbf{n}$.
The energy of the two-dineutron system in the finite volume, $E^{(L)}$, is related to the relative momentum $p$ by
\begin{align}
    E^{(L)} =& \frac{p^2}{2\mu} - 2 B_{2n} - 2 \Delta B_{2n}^{(L)} \left[ \sum_{\mathbf{k}} \frac{1}{(\mathbf{k}^2-\eta)^2} \right]^{-1} \notag \\
    & \times \sum_{\mathbf{k}} \frac{\sum_{i=1}^{3} \cos(2\pi\alpha k_i)}{3(\mathbf{k}^2-\eta)^2}.
    \label{eqn:E-momentum-01}
\end{align}
Here, $\mu$ is the reduced mass of the two dineutrons, $B_{2n}$ is the dineutron binding energy, $\alpha = 1/2$ for the dineutron, and $\Delta B_{2n}^{(L)} = B_{2n}^{(L)} - B_{2n}$ the finite volume correction.
The last term is a topological factor that accounts for the finite-volume momentum-dependent effects \cite{Bour:2011ef},
where the summation over $\mathbf{k}$ includes all integer vectors.

\section{Results and discussion}

We employ a lattice spacing of $a = 1.32$ fm, with $L$ ranging from 8 to 23, corresponding to physical box sizes from approximately 10~fm to 30~fm.
For comparison, we adopted an interaction with SU(4) symmetry to fit the $^1$S$_0$ scattering phase shifts. Details regarding its specific form and parameters are provided in Supplementary Material.

\begin{figure}[!htbp]
  \includegraphics[width=0.4\textwidth]{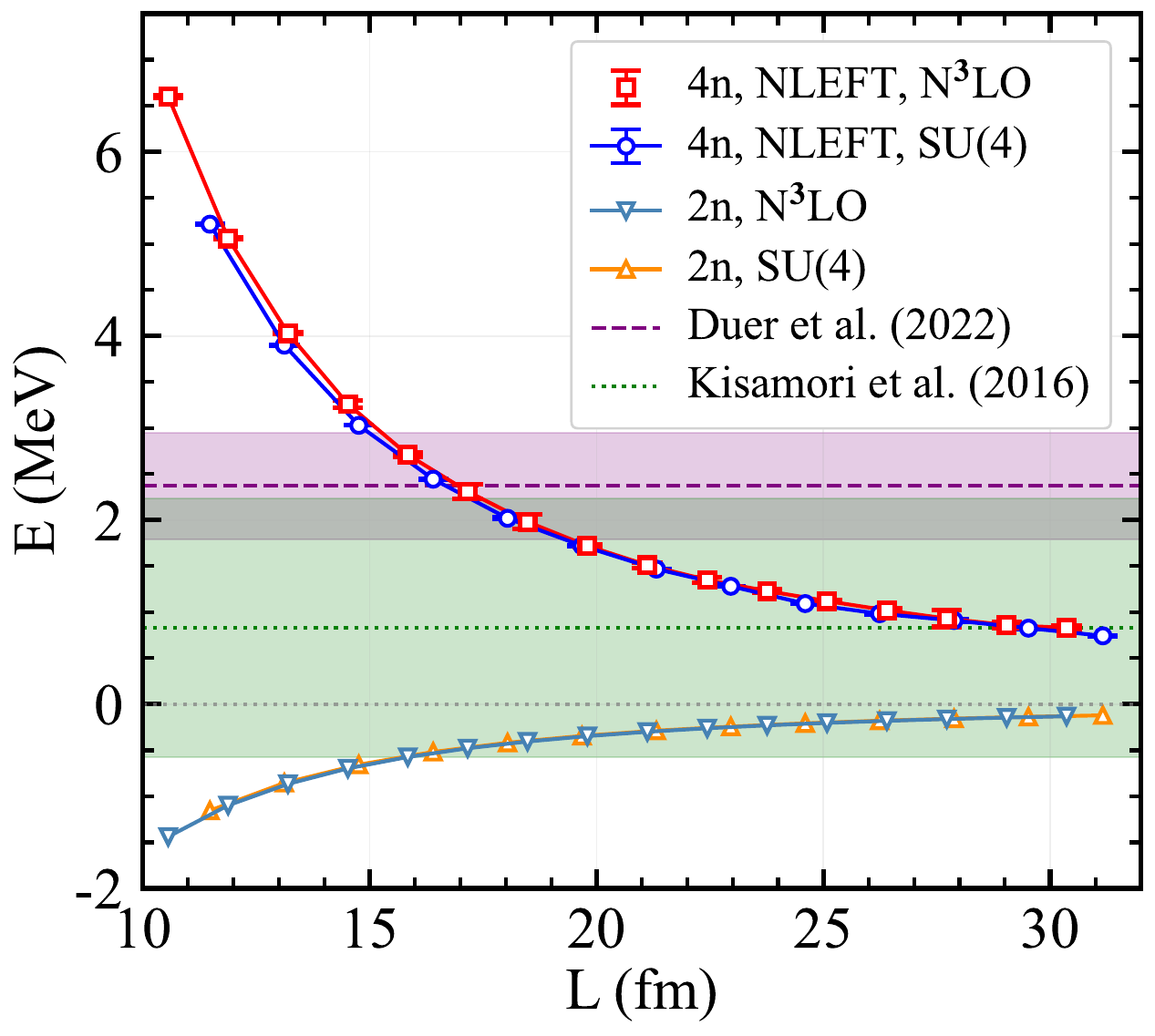}
  \caption{Ground-state energies of the $4n$ system calculated using NLEFT with both N$^3$LO and SU(4) interactions, and the $2n$ system calculated using the Lanczos eigenvector method with the SU(4) interaction, as functions of the lattice size $L$. The error bars of NLEFT results indicate the uncertainty from the extrapolation~\cite{SM}.}
  \label{fig:energy}
\end{figure}

\nocite{Zhang:2007hn}
\nocite{Lu:2018bat}
\nocite{Elhatisari:2016owd}
\nocite{Maier:1980}
\nocite{Savitzky:1964}
\nocite{Luscher:1985dn}
\nocite{Beane:2006mx}
\nocite{Stoks:1993tb}
\nocite{Lanczos:1950zz}
\nocite{Hazi:1970}
\nocite{Bedaque:1997qi}
\nocite{Bedaque:1998mb}
\nocite{Gabbiani:1999yv}
\nocite{Bedaque:2002yg}

Fig.~\ref{fig:energy} shows the calculated energy of the tetraneutron system at different $L$ values, using both the full N$^3$LO interaction ($a = 1.32$~fm) and the simple SU(4) interaction ($a = 1.64$~fm).
We also indicate candidate resonance energies from experimental studies: $E = 2.37 \pm 0.38\text{(stat.)} \pm 0.44\text{(sys.)}\,$MeV from Ref.~\cite{Duer:2022ehf} and $E = 0.83 \pm 0.65\text{(stat.)} \pm 1.25\text{(sys.)}\,$MeV from Ref.~\cite{Kisamori:2016jie}.
For comparison, the energies of the $2n$ system at different $L$ values using the SU(4) and N$^3$LO interactions are computed via the Lanczos eigenvector method~\cite{Lanczos:1950zz}, which eliminates Monte Carlo and extrapolation uncertainties.

Both the N$^3$LO and SU(4) interactions yield similar results for the $4n$ system, showing strongly repulsive behavior at small $L$ due to Pauli blocking, with energies gradually decreasing as $L$ increases.
This consistency confirms earlier findings~\cite{Lazauskas:2005ig,Hiyama:2016nwn,Gandolfi:2016bth,Shirokov:2016ywq,Fossez:2016dch,Deltuva:2018xoa,Li:2019pmg,Marques:2021mqf} that the tetraneutron system is insensitive to details of the nuclear interaction as long as it is realistic.
The smooth, continuous decrease of the $4n$ energy lacks the characteristic plateau formation expected for resonances in finite-volume methods~\cite{Hazi:1970,Zhang:2007hn}, suggesting a behavior typical of a non-bound system that does not form a resonance.
Furthermore, we have examined the energy derivatives and find no evidence of resonant behavior in the tetraneutron system~\cite{SM}.
Calculations at large box sizes are computationally demanding and yield larger Monte Carlo uncertainties. Nevertheless, we extend our N3LO calculations up to $L\simeq 30$~fm, reaching energies down to $E\simeq 0.83$~MeV, i.e.\ within the range of experimental candidate tetraneutron resonance energies. In the $L=25$--$30$~fm window the energy decreases monotonically and does not become volume independent within uncertainties, so no plateau is observed.

Recent studies of the $4n$ energy distribution using the $^8$He($p,p^4$He)$4n$ reaction model suggest that the sharp low-energy peak observed by~\cite{Duer:2022ehf} can be explained by dineutron-dineutron correlations~\cite{Lazauskas:2022mvq}. 
In a finite volume, the $2n$ subsystem becomes quasibound in the sense that the lowest $S$-wave level is shifted to negative
energy relative to the two-neutron threshold, even though the infinite-volume $nn$ system corresponds to a virtual state. This motivates us to study $2n$--$2n$ scattering phase shifts using L\"uscher's finite-volume method, following
Ref.~\cite{Elhatisari:2016hui}. In this context, we employ a finite-volume dineutron approximation, which means that the two-neutron
subsystem is treated as an effective weakly bound dimer inside the box, characterized by a large scattering length.

\begin{figure}[!htbp]
  \includegraphics[width=0.43\textwidth]{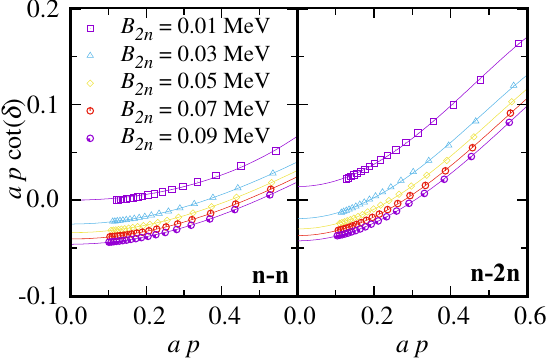}
  \caption{Neutron--neutron ($n$--$n$) and neutron--dineutron ($n$--$2n$) scattering phase shifts calculated using L\"uscher's finite-volume method with the SU(4) interaction at a lattice spacing $a = 1.64\,$fm. Points represent lattice data for different $B_{2n}$ values, and lines show fits based on the effective range expansion.}
  \label{fig:phase-n}
\end{figure}

Before turning to $2n$--$2n$ scattering, we validate this finite-volume strategy by studying $n$--$n$ and $n$--$2n$ scattering. Since at low energies and large separations the interaction details are then suppressed, we consider the zero-range leading interaction here. We tune the coupling constant of the zero-range neutron-neutron interaction so that the two-neutron subsystem is close to the
threshold, and we parameterize this proximity by an effective ``binding'' scale $B_{2n}$ spanning $0.01$ to $0.09$~MeV.
We choose this range to explore shallow finite-volume dimers 
close to the $nn$ threshold. The lower end $B_{2n}=0.01$~MeV deliberately
probes the extreme near-threshold regime, where finite-volume and 
lattice-spacing effects are amplified (see discussion below), while the 
upper end $B_{2n}=0.09$~MeV remains sufficiently close to threshold to 
stay in the large-scattering-length regime but provides a practical 
range for systematic checks.

We then use the resulting couplings in the two-neutron and three-neutron systems and compute the finite-volume scattering
energy levels, from which we extract the phase shifts using L\"uscher's method, as shown in Fig.~\ref{fig:phase-n}. Here we
define all quantities in lattice units, i.e., physical units multiplied by the corresponding power of $a$.

The lattice data (points) and effective-range-expansion fits (lines) show that $p\cot\delta$ versus $p$ collapses onto the expected universal trend for all systems except at $B_{2n}=0.01$~MeV.
This behavior is related to the fact that for short-range interactions with a large two-body scattering length,
low-energy observables become insensitive to microscopic details of the interaction and are governed
instead by universal large-scattering-length physics associated with the unitarity (conformal) fixed
point~\cite{Braaten:2004rn}. In this limit, dimensionless combinations such as $a\,p\cot\delta$ versus $a\,p$ approach
universal functions, and ratios of scattering lengths (e.g., $a_{n\text{-}2n}/a_{n\text{-}n}$) tend toward
constants, up to subleading range and regulator effects.

The behavior at $B_{2n}=0.01$~MeV reflects that we are extremely close to the
threshold regime where the intrinsic two-neutron length scale becomes very large, and small finite-volume and lattice-spacing effects can be amplified and even shift the two-body pole between a shallow bound state and a virtual state. Consistent with this interpretation,
for all lattice spacings examined the neutron--neutron scattering length at $B_{2n}=0.01$~MeV becomes
large and negative, indicating a virtual state, and the neutron--dineutron system shows a similarly large and negative scattering length.

Using the effective-range expansion, we extract the ratio of neutron-dineutron to neutron-neutron scattering lengths, $a_{n\text{-}2n}/a_{n\text{-}n}$. The results, summarized in Table~\ref{tab:inbetween}, show a nearly constant ratio for most cases,
in agreement with lattice calculations $1.176(6)$~\cite{Elhatisari:2016hui} and semianalytic continuum calculations $1.1791(2)$~\cite{
Bedaque:1997qi,Bedaque:1998mb,Gabbiani:1999yv,Bedaque:2002yg}.
We have verified that the results are robust against variations in lattice spacing from
$a=1.32$ to $1.97$~fm~\cite{SM}.

Since L\"uscher's relation is derived for elastic scattering, another important consistency check is that the finite-volume levels used to extract $n$--$2n$ phase shifts are not significantly affected by the opening of the breakup channel ($n+2n\to n+n+n$). In the finite-volume dimer picture, the relevant inelastic scale is set by the separation energy of the quasibound $2n$ level, which we characterize by $B_{2n}$. This corresponds to a characteristic momentum scale $\sqrt{mB_{2n}}$, with $m$ the nucleon mass. For our weakest binding this scale is small; nevertheless, even when the extracted momenta are larger than this estimate, we do not observe systematic departures
from the effective-range-expansion description or anomalous volume dependence in $p\cot\delta$. This indicates that breakup
(inelastic) effects are numerically negligible for the particular states and volumes analyzed here, and supports the use of
the finite-volume dineutron approximation as an intermediate step toward $2n$--$2n$ scattering.

\begin{table}[!htp]
    \centering
    \caption{Ratio of neutron--dineutron to neutron--neutron scattering lengths for different dineutron binding energies $B_{2n}$ at a lattice spacing $a = 1.64$~fm. The column of uncertainty indicates the statistical uncertainty from the fits to the effective range expansion.}
    \begin{tabular*}{0.5\textwidth}{@{\extracolsep{\fill}}lcc@{}}
    \hline
    \hline
    $B_{2n}$ & $a_{n-2n}/a_{n-n} $ & Uncertainty \\
    \hline
    $0.01$ & $3.35\times10^{-7}$ & $2.70$ \\
    $0.03$ & $1.29$ & $0.99$ \\
    $0.05$ & $1.13$ & $0.27$ \\
    $0.07$ & $1.09$ & $0.11$ \\
    $0.09$ & $1.08$ & $0.06$ \\
    \hline
    \hline
    \end{tabular*}
    \label{tab:inbetween}
\end{table}

We now turn to the extraction of the $S$-wave $2n$--$2n$ phase shifts from our finite-volume spectra.
For the realistic N$^{3}$LO chiral EFT Hamiltonian of Ref.~\cite{Elhatisari:2022zrb}, the infinite-volume
two-neutron system is not bound, but it corresponds to a virtual state.
Nevertheless, in the periodic boxes used here the lowest $nn$ $S$-wave level is shifted below the
two-neutron threshold and can be treated as a shallow finite-volume dimer (a quasibound dineutron).
This provides a practical intermediate description in which the four-neutron finite-volume levels are interpreted
as two composite dimers in relative motion, and L\"uscher's relation can be used to map the discrete energies
to elastic $2n$--$2n$ phase shifts.

For each box size $L$ we compute the four-neutron energy $E_{4n}^{(L)}$ and the two-neutron energy $E_{2n}^{(L)}$ in the same periodic volume.
The latter defines the effective finite-volume ``binding'' of the quasibound $2n$ level,
$B_{2n}^{(L)}\equiv -E_{2n}^{(L)}$.
To extract the relative momentum $p$ of the two-dineutron system we invert the finite-volume
dispersion relation for composite dineutrons [Eq.~(\ref{eqn:E-momentum-01})].
This inversion requires as input the infinite-volume dineutron binding energy $B_{2n}$.
For the N$^{3}$LO interaction this scale tends to zero as $L\!\to\!\infty$ and cannot be determined directly
from our largest volumes.
We therefore treat $B_{2n}$ as an auxiliary near-threshold parameter and scan representative values
$B_{2n}=0.01$--$0.09$~MeV (all below the smallest $B_{2n}^{(L)}$ encountered at our largest $L$),
using the resulting spread as an estimate of the residual systematic uncertainty associated with employing
the finite-volume dineutron picture when the underlying $nn$ pole is virtual.

\begin{figure}[!htbp]
  \includegraphics[width=0.43\textwidth]{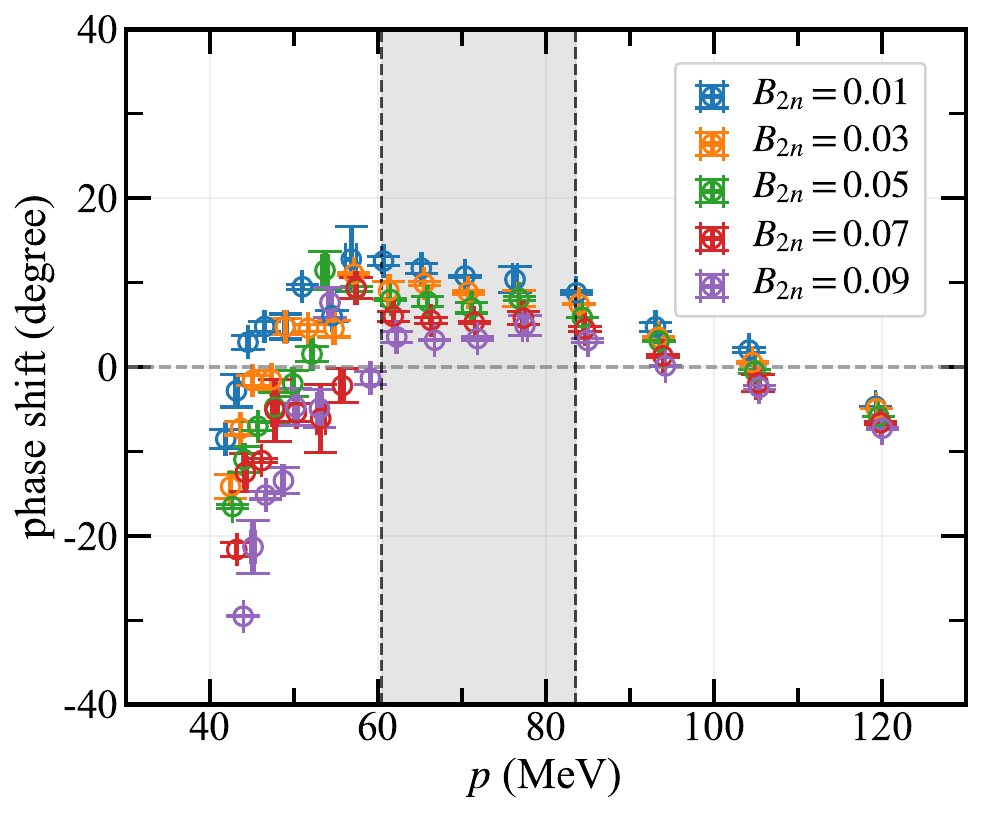}
  \caption{Dineutron--dineutron $S$-wave scattering phase shift $\delta$ as a function of relative momentum $p$ extracted from finite-volume $4n$ and $2n$ energies using L\"uscher's method and the composite-dimer finite-volume relation in Eq.~(\ref{eqn:E-momentum-01}), for the N$^{3}$LO chiral EFT interaction of Ref.~\cite{Elhatisari:2022zrb}. Different symbols correspond to the scanned near-threshold values of $B_{2n}$ (in MeV), and the shaded band indicates the momentum range associated with $L\simeq 20$--$15$~fm.}
  \label{fig:phase}
\end{figure}
The resulting $2n$--$2n$ $S$-wave phase shifts are shown in Fig.~\ref{fig:phase}. At the smallest extracted relative momenta (largest boxes), the phase shifts show the largest sensitivity
to the auxiliary near-threshold input $B_{2n}$ used in the inversion of Eq.~(\ref{eqn:E-momentum-01}). 
While some of the $B_{2n}$ values yield slightly negative $\delta(p)$ in this regime, the low-momentum trend
systematically bends toward $\delta(p)\to 0$ as $B_{2n}\to 0$, as expected for short-range interactions. We therefore do not draw a firm conclusion from the sign of the lowest-momentum points in Fig.~\ref{fig:phase}. As the momentum increases, $\delta(p)$ rises, becomes positive, and exhibits a shallow maximum
of order $10^\circ$ around $p\simeq 60$~MeV, before decreasing again and turning negative at the
highest momenta shown.
The dependence on the scanned $B_{2n}$ values is mild over most of the range and is most pronounced
only at the smallest momenta, where the dineutron size is largest and finite-volume effects are maximally amplified.

In the intermediate window $p\simeq 55$--$100$~MeV, $\delta(p)$ displays a clear nonmonotonic structure,
indicating a weak attractive feature.
However, we do not observe the characteristic rapid rise through $90^\circ$ that would signal a narrow
elastic resonance.
The shaded band in Fig.~\ref{fig:phase} highlights $p\simeq 60$--$84$~MeV, which corresponds to
$L\simeq 20$--$15$~fm and (for the same volumes) to confined $4n$ energies in the range
$E_{4n}\simeq 1.7$--$3.3$~MeV (cf.~Fig.~1), overlapping the experimentally reported correlated $4n$ peak
at $2.37\pm0.38\text{(stat.)}\pm0.44\text{(sys.)}$~MeV.

Our results are qualitatively consistent with previous investigations of $2n$--$2n$ scattering phase shifts using rigorous solutions of FY and AGS equations~\cite{Deltuva:2019mnv}, which found that at physical interaction strengths, a bound tetraneutron state evolves into a virtual state rather than a resonance. While we identify a weak nonmonotonic attraction at intermediate momenta in the current framework, the phase shift remains well below $90^\circ$, supporting the conclusion that no observable narrow resonance exists at the physical point.

\section{Summary}

We have investigated the tetraneutron system confined in a cubic box using nuclear lattice effective field theory, with box sizes up to $L = 30$~fm.
Using the N$^{3}$LO chiral EFT interaction, the confined $4n$ energy decreases gradually with increasing $L$ and remains concave over the explored volumes, showing no plateau that would be characteristic of a narrow resonance.
For comparison, we also repeat the confined-spectrum calculation with an SU(4)-symmetric interaction fitted to the $^1$S$_0$ phase shifts; the SU(4) results are consistent with the N$^{3}$LO trend and are reported in~\cite{SM}.

To interpret possible $4n$ correlations in terms of two-dineutron dynamics, we employ the finite-volume dineutron approximation, where the lowest $nn$ $S$-wave level in a periodic box is quasibound even though the infinite-volume $nn$ system corresponds to a virtual state.
Within this framework, we validate the L\"uscher-based extraction by studying $n$--$n$ and $n$--$2n$ scattering.
The ratio of scattering lengths $a_{n-2n}/a_{n-n}$ exhibits the expected universal behavior for near-threshold systems, with deviations only when $B_{2n}\approx 0.01$~MeV, where the proximity to the virtual-state regime amplifies finite-volume and lattice-spacing effects.

We then extract the $2n$--$2n$ $S$-wave phase shifts from finite-volume $4n$ and $2n$ energies using L\"uscher's method together with the composite-dineutron finite-volume relation.
At the smallest momenta, the extracted phase shifts show the largest 
sensitivity to $B_{2n}$. While for some of $B_{2n}$ values we observe slightly negative 
$\delta(p)$ in this regime, the low-momentum trend bends toward $\delta(p)\to 0$ as 
$B_{2n}\to 0$, and we therefore do not draw a firm conclusion from the 
sign of the lowest-momentum points.
In the intermediate window $p\simeq 55$--$100$~MeV, $\delta(p)$ becomes positive and shows a shallow nonmonotonic structure, with a maximum of order $10^\circ$ around $p\simeq 60$--$84$~MeV, but without a rapid rise through $90^\circ$ typical of a narrow elastic resonance.
This momentum band corresponds to box sizes $L\simeq 20$--$15$~fm and confined $4n$ energies of $1.7$--$3.3$~MeV, which lie close to the experimentally reported correlated $4n$ peak at $2.37\pm0.38\text{(stat.)}\pm0.44\text{(sys.)}$~MeV.

In the future, we plan to extend the NLEFT framework~\cite{Elhatisari:2015iga} to investigate the $^8$He($p,p^4$He)$4n$ reaction and understand the correlated structures observed. As a first step in this direction, NLEFT has recently been used to quantify four-neutron correlation patterns in $^{7}$H and $^{8}$He, motivated by quasifree knockout measurements~\cite{Zhang:2025uin}.


\section{Acknowledgments}
We are grateful for discussions with Kouichi Hagino, Hiroyuki Sagawa, Furong Xu, and the members of the NLEFT Collaboration.
This work is supported by National Natural Science Foundation of China under Grant No. 12435007 and Fundamental and Interdisciplinary Disciplines Breakthrough Plan of the Ministry of Education of China. The work of S.E. is supported in part by the Scientific and Technological Research Council of Turkey (TUBITAK Project No. 123F464). The work of U.-G. M. was supported in by the European Research Council (ERC) under the
European Union’s Horizon 2020 research and innovation program (ERC AdG EXOTIC, Grant Agreement No. 101018170), by the CAS President’s International Fellowship Initiative (PIFI) (Grant No. 2025PD0022). The work of Y. K. was supported in part by the Institute for Basic Science (IBS-R031-D1). L.W. gratefully acknowledge the Computational resources provided by the HPC platform of Beihang University, the National Supercomputing Center of Korea with supercomputing resources including technical support (KSC-2024-CHA-0001, KSC-2025-CHA-0004). S.S. and U.-G. M. gratefully acknowledge the Gauss Centre for Supercomputing e.V. for funding this project by providing computing time on the GCS Supercomputer JUWELS at the J\"ulich Supercomputing Centre (JSC).

\emph{Data availability}---The data that support the findings of this article are openly available~\cite{Wu2026}.


\bibliography{bref-nleft}


\clearpage

\beginsupplement

\begin{bibunit}[apsrev4-2]
\input{sm-4nnleft}
\putbib[bref-nleft]
\end{bibunit}

\end{document}

%% file: sm-4nnleft.tex






\begin{onecolumngrid}
\renewcommand{\thefigure}{S\arabic{figure}}
\renewcommand{\thetable}{S\arabic{table}}
\renewcommand{\theequation}{S\arabic{equation}}
\setcounter{figure}{0}
\setcounter{equation}{0}
\setcounter{table}{0}

\section*{Supplementary Material}

\subsection{Energy extrapolation}
For each $L$, the energy is computed at several projection times $\tau$ and extrapolated to the infinite-time limit $E_{\infty}^{(L)}$ using~\cite{Shen:2024qzi}
\begin{equation}\label{eq:extrapolation}
    E^{(L)}(\tau)=\frac{E_{\infty}^{(L)}+\left(E_{\infty}^{(L)}+d^{(L)}\right) c^{(L)} e^{-d^{(L)} \tau}}{1+c^{(L)} e^{-d^{(L)} \tau}}
\end{equation}
where $E_{\infty}^{(L)}$, $d^{(L)}$, and $c^{(L)}$ are fitting parameters.
We find that the dominant contribution to the interaction energy comes from the $^1$S$_0$ channel, consistent with previous studies~\cite{Lazauskas:2005ig}.
Other contributions, including the three-nucleon force, are negligibly smaller.
For comparison, we also employ an SU(4) symmetric interaction~\cite{Lu:2018bat} fitted to the $^1$S$_0$ phase shift, with the form:
\begin{equation}\label{eq:}
    V_{\rm SU(4)} = \frac{1}{2} C_2 \sum_{\mathbf{n}} \tilde{\rho}(\mathbf{n})^2.
\end{equation}
The density operator is defined as
\begin{equation}\label{eq:}
    \tilde{\rho}(\mathbf{n}) = \sum_i \tilde{a}_i^\dagger(\mathbf{n}) \tilde{a}_i(\mathbf{n}) + s_{\rm L} \sum_{|\mathbf{n}'-\mathbf{n}|=1} \sum_i \tilde{a}_i^\dagger(\mathbf{n}') \tilde{a}_i(\mathbf{n}'),
\end{equation}
where $i$ denotes the spin index, $s_{\rm L}$ is a local smearing parameter, and $\tilde{a}_i^\dagger(\mathbf{n})$ represents the nonlocally smeared creation operator~\cite{Elhatisari:2016owd}:
\begin{equation}\label{eq:}
    \tilde{a}_i^\dagger(\mathbf{n}) = {a}_i^\dagger(\mathbf{n}) + s_{\rm NL} \sum_{|\mathbf{n}'-\mathbf{n}|=1} {a}_i^\dagger(\mathbf{n}').
\end{equation}
Given the dilute nature of the system, we calculate this SU(4) interaction at a larger lattice spacing $a = 1.64$~fm for comparison.
The fitted parameters are $s_{\rm NL} = 0.1$, $s_{\rm L} = 7.3918 \times 10^{-2}$, and $C_2 = -8.4272 \times 10^{-6}\,$MeV$^{-2}$.

By employing Monte Carlo sampling to calculate the energies of the tetraneutron system at different values of \(\tau\) for each lattice size \(L\), and using Eq.~(\ref{eq:extrapolation}) to extrapolate to infinite projection time, the extrapolation results for both the N$^3$LO and SU(4) interactions are obtained, as shown in Fig.~\ref{fig:N3LO-extrapolation}  and Fig.~\ref{fig:SU4-extrapolation}, respectively.


\begin{figure}[!htbp]
\centering
\includegraphics[width=1.0\textwidth]{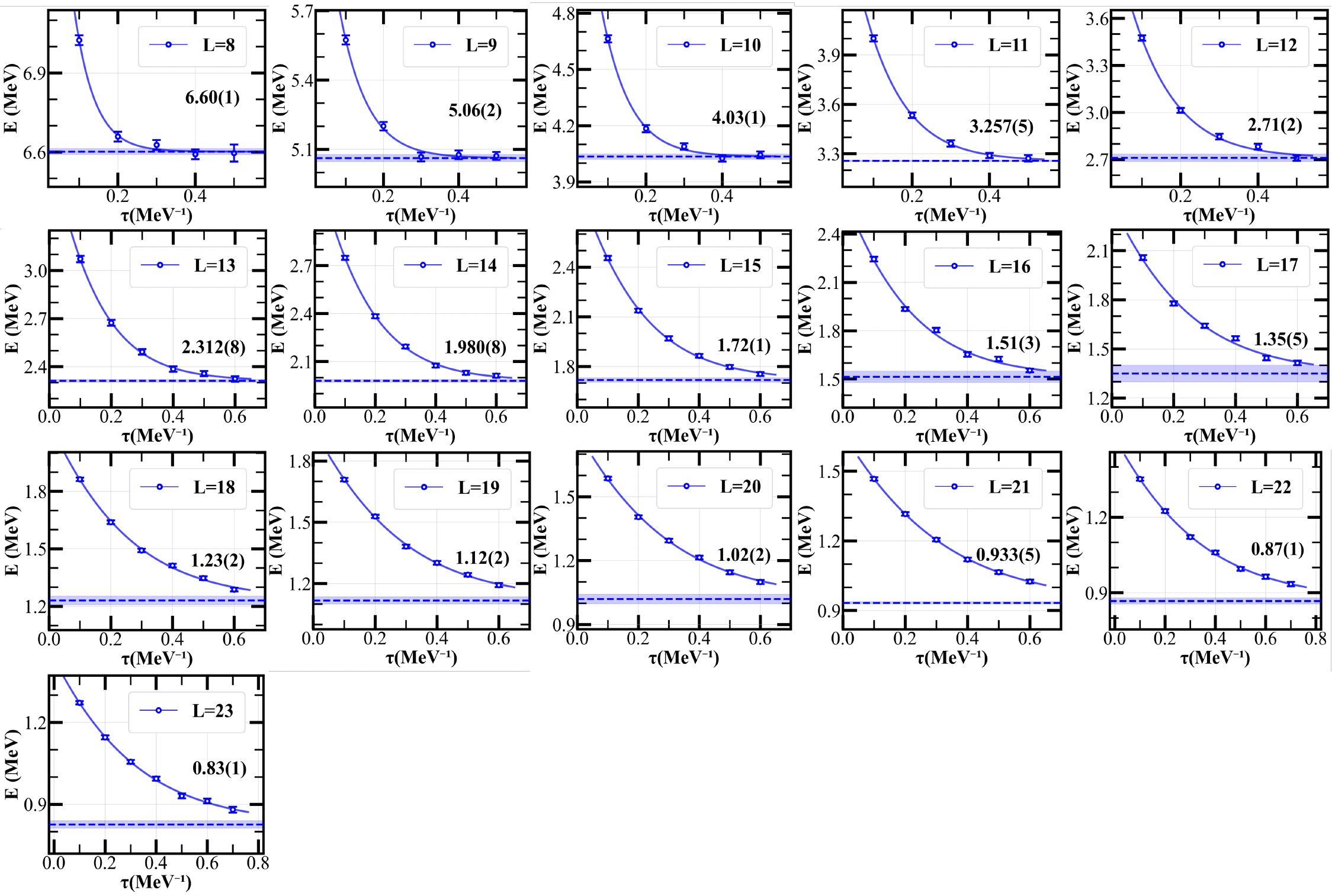}
\caption{The energy extrapolation of the four-neutron system with the N$^3$LO interaction at various lattice sizes $L$.}
\label{fig:N3LO-extrapolation}
\end{figure}


\begin{figure}[!htbp]
\centering
\includegraphics[width=1.0\textwidth]{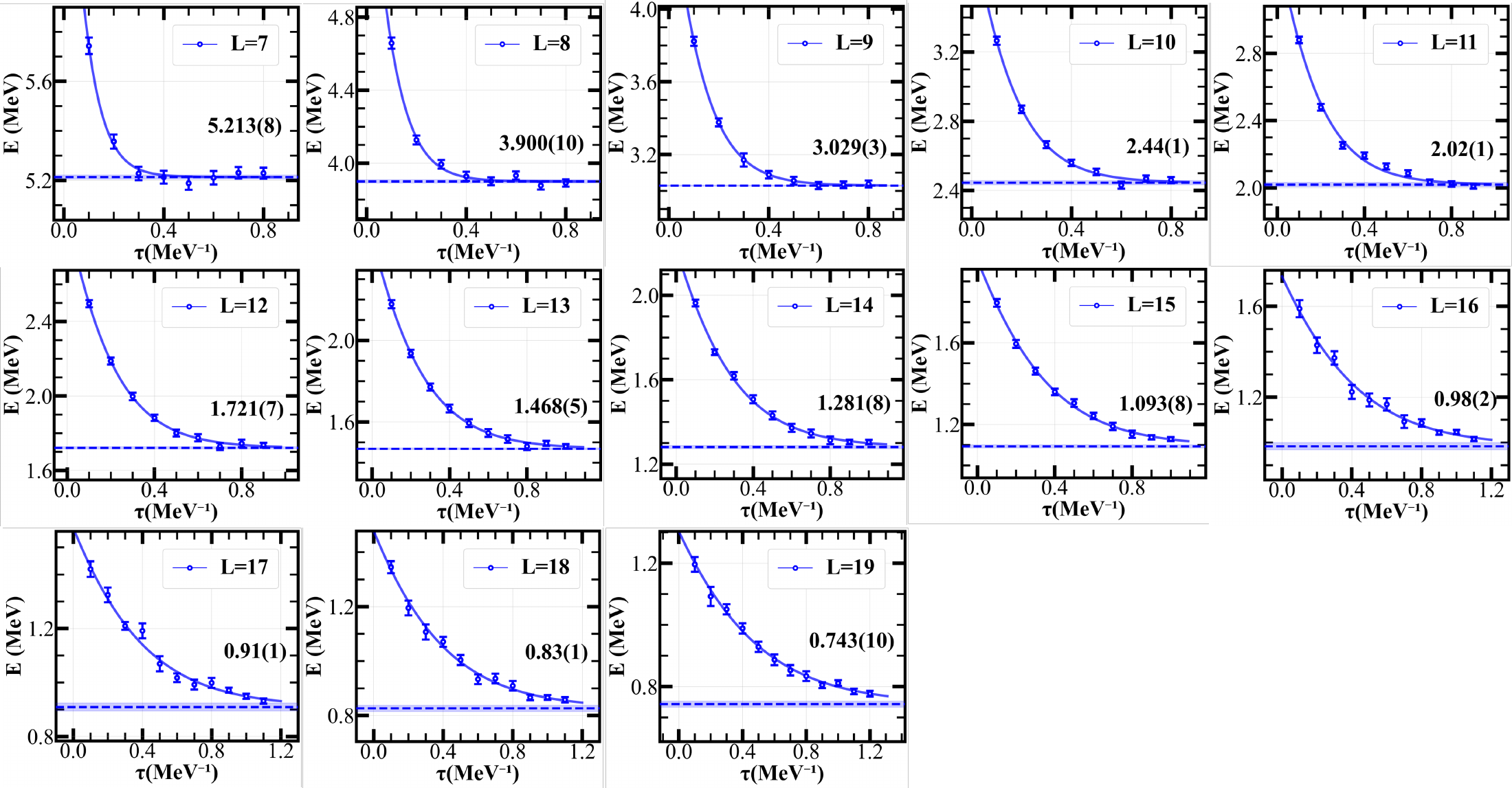}
\caption{The energy extrapolation of the four-neutron system with the SU(4) interaction at various lattice sizes $L$.}
\label{fig:SU4-extrapolation}
\end{figure}

\subsection{Energy derivative}

Following Refs.~\cite{Maier:1980,Zhang:2007hn}, we compute the derivatives of the energy with respect to the box size $L$, as shown in Fig.~\ref{fig:nabla}.
The numerical derivatives, particularly the second derivative for the N$^3$LO interaction, exhibit relatively large uncertainties due to Monte Carlo sampling and projection-time extrapolation.
For comparison, we also show results smoothed using the Savitzky-Golay filter~\cite{Savitzky:1964}.
In neither case do we observe the clear signature of a resonance, which would require $\partial^2 E/\partial L^2$ to cross zero.
The energy curve remains concave and does not transition to convex behavior.

\begin{figure}[!htbp]
  \includegraphics[width=1.0\textwidth]{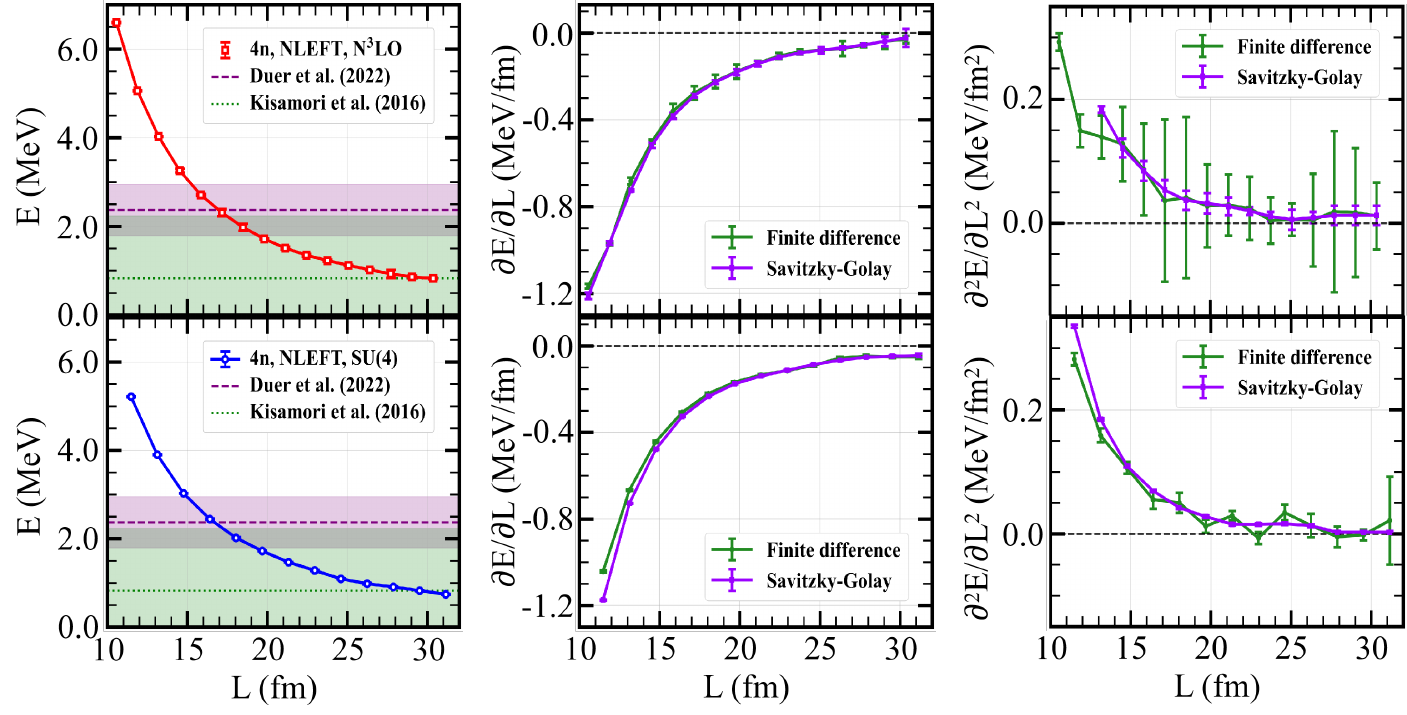}
  \caption{Energy, energy gradient, and second derivative of the four-neutron system for the N$^3$LO and SU(4) interactions as functions of the lattice size $L$.}
  \label{fig:nabla}
\end{figure}

\subsection{finite-volume extrapolation for dineutron}

Figure~\ref{fig:e2n-L} shows the finite-volume dependence of the two-neutron ground-state energy computed with the SU(4) interaction. To assess the infinite-volume limit we use an empirical exponential form,
$E(L)=E_\infty + c\,e^{-\delta E\,L}/L$, and systematic large-$L$ expansions in $1/L$. The latter follow from the threshold expansion of L\"uscher’s finite-volume quantization condition for short-range interactions, which predicts power-law finite-volume shifts of the lowest two-body
level (with leading behavior $\propto 1/L^{3}$)~\cite{Luscher:1985dn,Luscher:1990ux,Beane:2006mx}.
Including higher order terms yields an infinite-volume energy consistent with zero,
$E_\infty \approx 0$.

\begin{figure}[!htbp]
  \includegraphics[width=0.5\textwidth]{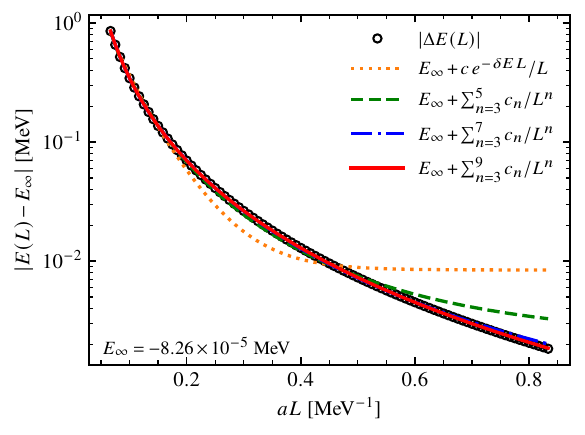}
  \caption{Dineutron energy as a function of the cubic box length $L$ for the SU(4) interaction. The curves correspond to infinite-volume extrapolations using different numbers of terms in the fit function.}
  \label{fig:e2n-L}
\end{figure}

\subsection{Phase shift}

The neutron-neutron $^1$S$_0$ scattering phase shift $\delta$, extracted from finite-volume two-neutron energies via L\"uscher’s method, is shown in Fig.~\ref{fig:phase-nn}. Results obtained with the SU(4) and N$^3$LO interactions are compared with experimental data~\cite{Stoks:1993tb}.

\begin{figure}[!htbp]
  \includegraphics[width=0.5\textwidth]{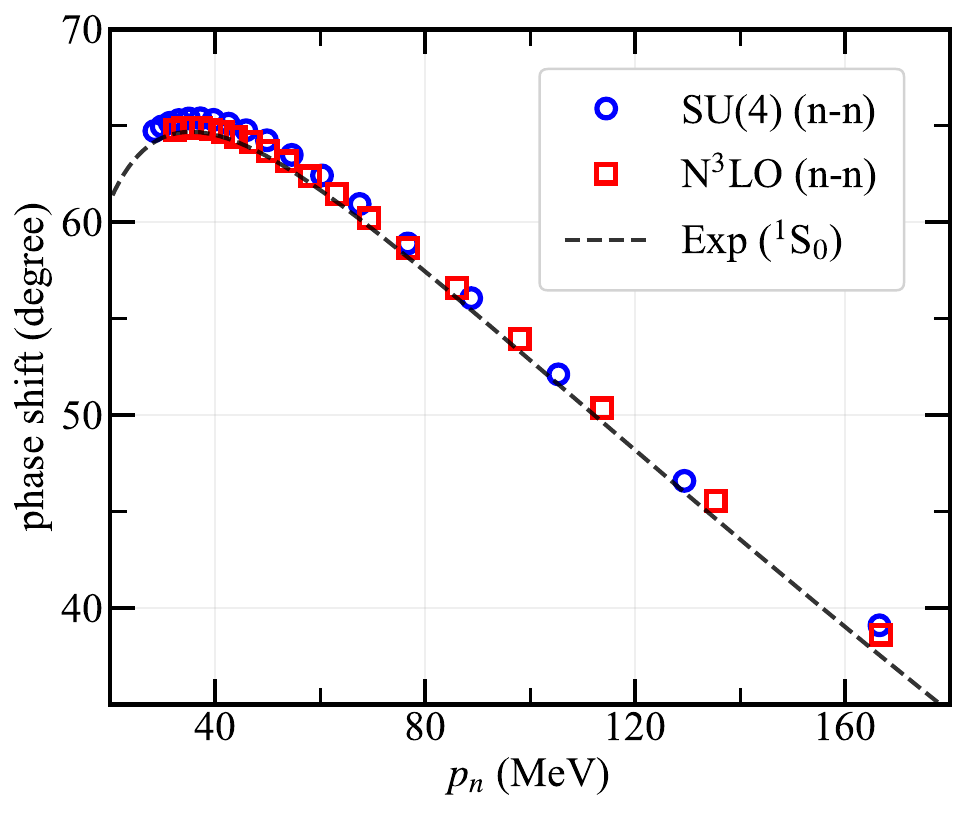}
  \caption{Neutron-neutron $^1$S$_0$ scattering phase shift $\delta$ as a function of relative momentum $p$. The points are calculated from finite-volume energies using Lüscher’s method for the SU(4) and N$^3$LO interactions, and are compared to the experimental analysis of Ref.~\cite{Stoks:1993tb}.}
  \label{fig:phase-nn}
\end{figure}

To examine whether the dineutron approximation can be used to study dineutron systems and scattering properties, we performed calculations of phase shifts for the simpler neutron-neutron and neutron-dineutron systems. The elastic two-body scattering phase shift is expanded via the effective range expansion as 
\begin{equation}\label{eq:scatter}
p \cot \delta = -\frac{1}{a_{n-n}} + \frac{1}{2} r_{n-n} p^{2} + \mathcal{O}(p^{4}),
\end{equation}
where \(p\) is the relative momentum of the two bodies, \(a_{n-n}\) is the neutron-neutron scattering length, and \(r_{n-n}\) is the neutron-neutron effective range. Using Eq.~(\ref{eq:scatter}) for extrapolation, the neutron-dineutron scattering length \(a_{n-2n}\) and effective range \(r_{n-2n}\) can also be determined by fitting the momentum and phase shift data at different \(L\).

To test the universality of the dineutron approximation under pure contact interactions, we varied the dineutron binding energy from 0.1~MeV down to nearly zero at lattice spacings \(a = 1.32,\;1.64,\;1.97\;\text{fm}\), computed the relation between the dimensionless quantity \(ap\cot\delta\) and \(ap\) as shown in Fig.~\ref{fig:d-n scattering}. Except for the system with a binding energy of 0.01~MeV, all results align with the expectations, indicating the validity of the dineutron approximation. 

Based on Eq.~(\ref{eq:scatter}), we calculated the neutron-dineutron scattering length \(a_{n-2n}\), neutron-neutron scattering length \(a_{n-n}\), and their ratio \(a_{n-2n}/a_{n-n}\) for different lattice spacings corresponding to a dineutron binding energy \(B_{2n} = 0.01~\text{ MeV}\). The results are listed in Table~\ref{tab:scattering_l}. Here, the values of \(a_{n-n}\) are large and negative across all lattice spacings, indicating a virtual state in the system, while the neutron-dineutron system also exhibits large negative scattering lengths. Additionally, Table~\ref{tab:ratios} presents the ratio of neutron-dineutron to neutron-neutron scattering lengths for dineutron binding energies ranging from \(B_{2n} = 0.01\) to \(0.1~\text{ MeV}\) at different lattice spacings. The results show that for lattice spacings \(a = 1.32, 1.64,\) and \(1.97~\text{ fm}\), the values of \(a_{n-2n}/a_{n-n}\) exhibiting a universal behavior independent of the lattice spacing.

\begin{figure}[!htbp]
\centering
\includegraphics[width=1.0\textwidth, height=12cm]{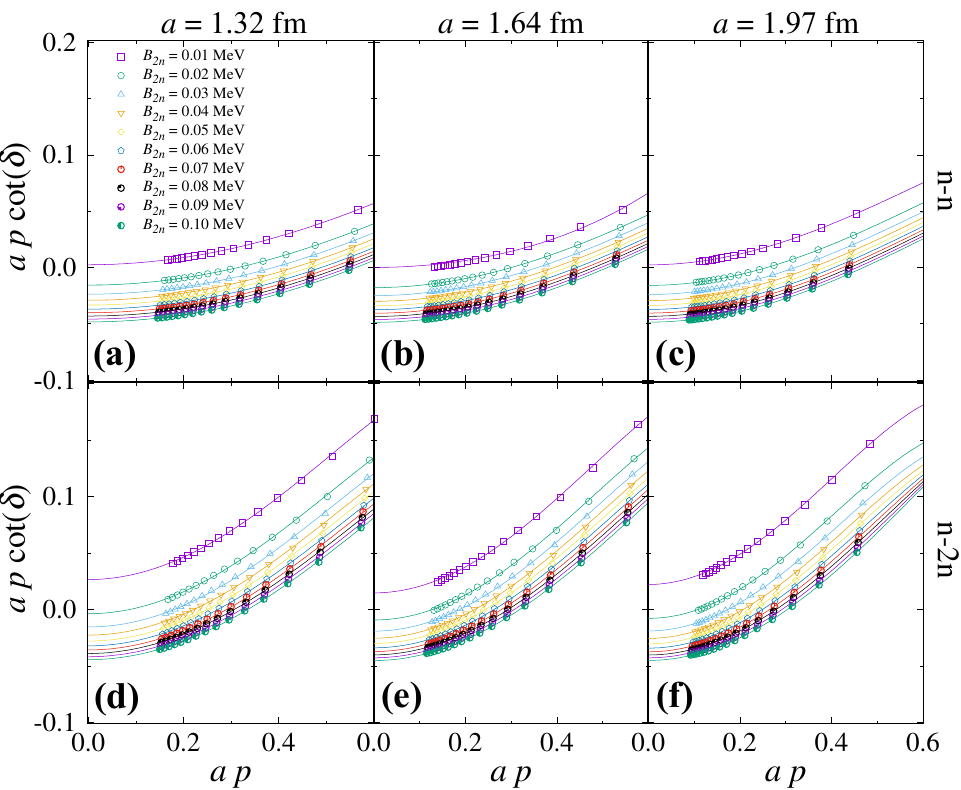}
\caption{Neutron-neutron (a-c) and neutron-dineutron (d-f) scattering phase shifts calculated using L\"uscher’s finite-volume method with the SU(4) interaction at lattice spacings $a = 1.32$ fm (a,d), $1.64$ fm (b,e), and $1.97$ fm (c,f). Points represent lattice data for different $B_{2n}$ values, and lines show fits based on the effective range expansion.}
\label{fig:d-n scattering}
\end{figure}

\begin{table}[!htp]
\centering
\caption{Scattering lengths for different values of $a$ values ($B_{2n}=0.01$~MeV). The column of uncertainty indicates the statistical uncertainty from the fits to the effective-range expansion.}
\begin{tabular}{l|cc|cc|cc}
\hline
\hline
\multirow{2}{*}{quantity} & \multicolumn{2}{c}{\(a=1.32\) fm} & \multicolumn{2}{c}{\(a=1.64\) fm} & \multicolumn{2}{c}{\(a=1.97\) fm} \\
\cline{2-7}
& value & uncertainty & value & uncertainty & value & uncertainty \\
\hline
$a_{n-n}$ & -418.952822 & 4.714405 & $-3.625460\times10^8$ & $1.057483\times10^{14}$ & -433.063197 & 6.848189 \\
$a_{n-2n}$ & -37.524722 & 0.948073 & -67.299539 & 2.624853 & -44.818149 & 0.847960 \\
$a_{n-2n}/a_{n-n}$ & 0.089568 & 0.002477 & $1.856303\times10^{-7}$ & 0.054145 & 0.103491 & 0.002552 \\
\hline
\hline
\end{tabular}
\label{tab:scattering_l}
\end{table}

\begin{table}[!htp]
    \centering
    \caption{The ratio of neutron-dineutron to neutron-neutron scattering lengths for different lattice spacings and dineutron binding energies. The column of uncertainty indicates the statistical uncertainty from the fits to the effective-range expansion.}
    \begin{tabular}{l|cc|cc|cc}
    \hline
    \hline
    \multirow{2}{*}{$B_{2n}$} & \multicolumn{2}{c}{\(a=1.32\) fm} & \multicolumn{2}{c}{\(a=1.64\) fm} & \multicolumn{2}{c}{\(a=1.97\) fm} \\
    \cline{2-7}
    & $a_{n-2n}/a_{n-n}$ & uncertainty & $a_{n-2n}/a_{n-n}$ & uncertainty & $a_{n-2n}/a_{n-n}$ & uncertainty \\
    \hline
    $0.01$ & $0.089568$ & $0.002477$ & $1.86\times10^{-7}$ & $0.054145$ & $0.103491$ & $0.002552$ \\
    $0.02$ & $4.370859$ & $0.739432$ & $1.955848$ & $0.104053$ & $2.072985$ & $0.093956$ \\
    $0.03$ & $1.547733$ & $0.054318$ & $1.318567$ & $0.028219$ & $1.276168$ & $0.018059$ \\
    $0.04$ & $1.293470$ & $0.027550$ & $1.190182$ & $0.015986$ & $1.155832$ & $0.009129$ \\
    $0.05$ & $1.200866$ & $0.018532$ & $1.138108$ & $0.010687$ & $1.113122$ & $0.005522$ \\
    $0.06$ & $1.154382$ & $0.013838$ & $1.111648$ & $0.007609$ & $1.094222$ & $0.003659$ \\
    $0.07$ & $1.127364$ & $0.010883$ & $1.096796$ & $0.005587$ & $1.085393$ & $0.002799$ \\
    $0.08$ & $1.110502$ & $0.008832$ & $1.088154$ & $0.004205$ & $1.081597$ & $0.002647$ \\
    $0.09$ & $1.099464$ & $0.007320$ & $1.083139$ & $0.003294$ & $1.080561$ & $0.002883$ \\
    $0.10$ & $1.092066$ & $0.006170$ & $1.080411$ & $0.002781$ & $1.081122$ & $0.003255$ \\
    \hline
    \hline
    \end{tabular}
    \label{tab:ratios}
\end{table}

\begin{figure}[!htbp]
  \includegraphics[width=0.5\textwidth]{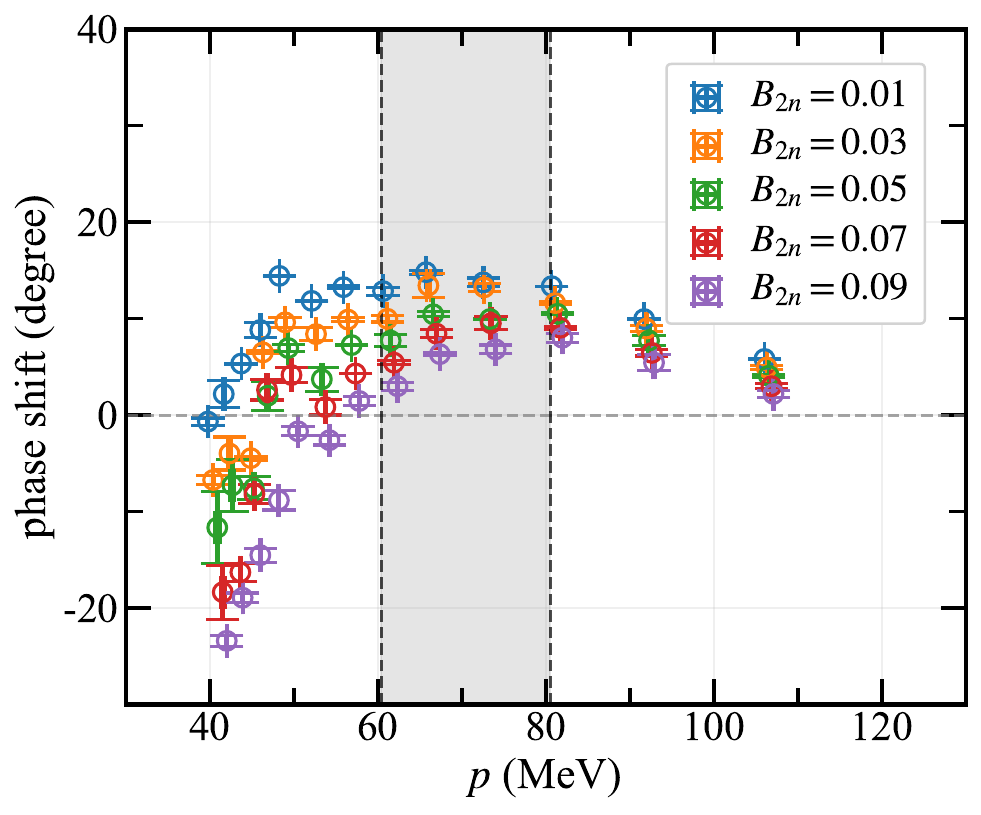}
  \caption{Dineutron-dineutron $S$-wave scattering phase shift $\delta$ as a function of relative momentum $p$ extracted from finite-volume $4n$ and $2n$ energies using L\"uscher's method and the composite-dimer finite-volume relation for the SU(4) interaction. Different symbols correspond to the scanned near-threshold values of $B_{2n}$ (in MeV), and the shaded band indicates the momentum range associated with $L\simeq 20$--$15$~fm.}
  \label{fig:phase-su(4)}
\end{figure}

We also calculated dineutron-dineutron $S$-wave scattering phase shifts for the SU(4) interaction, as shown in Fig.~\ref{fig:phase-su(4)}. In the momentum range 60-120~MeV, the phase shift also shows some attraction. Similar to the case with the N$^3$LO interaction, no rapid rise of the phase shift to $90^\circ$ is observed. The shaded region corresponds to $p\simeq$ 60–-80~MeV, which for a box size of $L\simeq$~20--15~fm yields a calculated tetraneutron energy range of $E\simeq1.7$–3.0 MeV.

\end{onecolumngrid}
